\input psfig.sty
\def\ktin{kT$_{\rm in}$}
\def\rin{R$_{\rm in}$}
\def\ktbb{kT$_{\rm BB}$}
\def\rbb{R$_{\rm BB}$}

\def\onee{1E1724-3045}

\def\la{\hbox{\rlap{\raise.3ex\hbox{$<$}}\lower.8ex\hbox{$\sim$}\ }}
\def\ga{\hbox{\rlap{\raise.3ex\hbox{$>$}}\lower.8ex\hbox{$\sim$}\ }}

\def\ergs{ergs s$^{-1}$}
\def\ergscm{ergs s$^{-1}$ cm$^{-2}$}

\def\nhv{H atoms cm$^{-2}$}
\def\nh{N$_{\rm H}$}
\def\kte{kT$_e$}
\def\ktw{kT$_W$}
\def\fbb{F$_{\rm BB}$}
\def\ft{F$_{\rm T}$}
\def\comptt{{\it Comptt}}
\def\ldbb{L$_{\rm DBB}$} 
\def\lbb{L$_{\rm BB}$} 
\def\lcomptt{L$_{\rm Comptt}$} 

\documentclass[11pt,psfig]{aa}
\usepackage{astron}

\begin{document}

   \thesaurus{06     
stars
              (08.14.1;
               08.02.3 
               10.07.2;
               13.25.3;
               13.25.5;
               13.07.2)} 
   \title{An ASCA X-ray Observation of \onee~in the globular cluster
   Terzan 2}


   \author{D. Barret\inst{1}, J. E. Grindlay\inst{2},
           I. M. Harrus\inst{2},  J. F. Olive\inst{1}}

   \offprints{D. Barret}

   \institute{Centre d'Etude Spatiale des Rayonnements, CNRS/UPS, 
9 Avenue du Colonel Roche, 31028 Toulouse Cedex 04, France 
(email: Didier.Barret@cesr.fr)
         \and
             Harvard Smithsonian Center for Astrophysics, 60 Garden 
Street,
             Cambridge, MA 02138, USA
             }

\date{Received ; accepted}

\titlerunning{ASCA observations of 1E1724-3045}
\authorrunning{Barret et al.}

  \maketitle

\begin{abstract} 
The bright persistent X-ray source and type I X-ray burster
\onee~located in the globular cluster Terzan 2 was observed by ASCA
for about 10 ksec on September 24-25th, 1995 while it was in its hard
state, with a luminosity of $\sim 6\times 10^{36}$ \ergs~in the 0.5-10
keV band (d=7.7 kpc). The ASCA spectrum is hard, and reveals the
presence of a soft component below 2-3 keV. When combined with non
simultaneous RXTE/HEXTE data, we show that the ASCA spectrum can be
adequately fitted by a Comptonization model for which the electron
temperature is $\sim 30$ keV, the optical depth is $\sim 3$ for a
spherical scattering cloud. The soft component carries off about 13\%,
and 35\% of the 0.1-200 keV luminosity for a blackbody and disk
blackbody fit respectively. It is more likely to arise from the
accretion disk, whereas the hard Comptonized component is generated in
a hot boundary layer interior to the disk.  The column density
measured towards the source (\nh $\sim 1.0\times10^{22}$ cm$^{-2}$) is
consistent with the value expected from the optical reddening of the
cluster. No emission lines were detected, and an upper limit of 30 eV
for the equivalent width of a 6.4 keV line ($\sigma=0.1$ keV) has been
derived.
\end{abstract}
\section{Introduction}
\onee~is a bright persistent Low-Mass X-ray Binary (LMXB) located in 
the globular cluster Terzan 2 \cite{tz2:grindlay80apjl}.  Terzan 2 is
a metal-rich globular cluster of the galactic bulge with a reddening
E(B-V)=1.57 \cite{tz2:ortolani97aa}.  Depending on the relation
R=$A_{\rm V}/{\rm E(B-V)}$, Ortolani et al.  \cite*{tz2:ortolani97aa}
estimated a distance of 7.7 kpc for the ``canonical'' value of 3.1 for
R \cite{savage79ara}, and d=5.3 kpc for R=3.6 \cite{grebel95aas}
respectively.

The type I X-ray bursts observed from \onee~indicate that the compact
object is a weakly magnetized neutron star
\cite{tz2:swank77apjl,tz2:grindlay80apjl}.  Analysis of these bursts, 
in particular the one observed by EINSTEIN that showed photospheric
radius expansion, and therefore likely reached the Eddington limit,
places the source at a distance of 7 kpc \cite{tanaka81}, consistent
with the estimate for R=3.1.  Recent monitoring of the source, by the
All Sky Monitor aboard the Rossi X-ray Timing Explorer
\cite{bradt93aas}, as well as snapshot observations performed earlier,
indicate that \onee~is weakly variable in X-rays (less than about a
factor of 3 on a few day time scale) with a mean flux level of 45
mCrab in the 2-12 keV range.

Prior to ASCA, EXOSAT, TTM and ROSAT X-ray observations ($\sim$ 1-20
keV) have shown that the source spectrum could be simply fit by a
power law of photon index $\sim 2.0-2.4$
\cite{tz2:mereghetti95aa,zand92thesis,verbunt95aa}.  At higher
energies, \onee~is one of the first neutron star systems from which
hard X-ray emission (E$\ga$35 keV) was discovered
\cite{tz2:barret91apjl}. 
 Additional observations of the source with SIGMA 
demonstrated that \onee~was a persistent, though
variable, hard X-ray source emitting at the mean level of 22 mCrab in
the 35-75 keV range
\cite{tz2:goldwurm93:2ndcgro,tz2:goldwurm94nat,tz2:goldwurm95asr}.
Fitting of the time-averaged SIGMA spectrum yields a photon index of
$3.0 \pm 0.3$ for a power law fit, thus suggesting the existence of a
cutoff or a break between the X-ray and hard X-ray bands \cite{bt97}.
\begin{table*}[t]
\scriptsize
\begin{center}
\begin{tabular}{llllcc}
\hline
Experiment & Date & E (keV) & Model & Flux & Ref.\\
\hline
\hline
EXOSAT ME+LE & 3/14/85 & 2--10 & PL: $\alpha_{\rm PL}=2.34^{+0.03}_{-0.04}$ & 
2.3 & 1 \\
& & & N$_{\rm H}=1.8^{+0.1}_{-0.1}$ & & \\
\hline
TTM & 10/88-02/92 & 2--28 & PL: $\alpha_{\rm PL}=2.1^{+0.2}_{-0.1}$ & 
7.9 & 2 \\
& & & N$_{\rm H}=4.0^{+1.0}_{-2.0}$ & & \\
\hline
ROSAT PSPC& 6/19-21/19 & 0.5--2.5 & PL: $\alpha_{\rm PL}=2.3$ & 3.1 & 3 \\ & &
& \nh=1.8 (assumed) & & \\
\hline
\hline
SIGMA & 03/90-10/93 & 35--75 & PL: $\alpha_{\rm PL}=3.0\pm 0.3$ & 1.8
& 4,5 \\
\hline
\hline
SAX & 08/17/96 & 0.1--100 & C+BB:\nh$=0.84^{+0.08}_{-0.07}$ & 15.8 & 6 \\
& & & \kte=$29^{+9}_{-4}$ & & \\
& & & $\tau=3.0^{+0.4}_{-0.5}$ & & \\
& & & \ktw$=1.1^{+0.10}_{-0.09}$ & & \\
& & & \ktbb$=0.6^{+0.05}_{-0.06}$ & & \\
& & & \rbb$=12.0^{+2.0}_{-2.0}$ & & \\
\hline
SAX & \ldots & \ldots & C+DBB:\nh=$1.22^{+0.04}_{-0.07}$ & 15.8 & 6 \\
& & & \kte=$27^{+11}_{-4}$ & & \\
& & & $\tau=3.2^{+0.5}_{-0.7}$ & & \\
& & & \ktw$=1.94^{+0.30}_{-0.16}$ & & \\
& & & \ktin$=1.4^{+0.15}_{-0.08}$ & & \\
& & & \rin $\sqrt{\cos\theta}=3.0^{+0.3}_{-0.4}$ & & \\
\hline
\hline
RXTE & 04-08/11/96 & 1--200 & C+DBB:\nh=$1.0$ (assumed) & 22.2 & 7 \\
& & & \kte=$33.8^{+14.0}_{-8.0}$ & & \\
& & & $\tau=2.4^{+0.4}_{-0.5}$ & & \\
& & & \ktw$=1.6^{+0.2}_{-0.3}$ & & \\
& & & \ktbb$=0.8^{+0.1}_{-0.1}$ & & \\
& & & \rbb$=7.0^{+1.7}_{-0.6}$ & & \\
\hline
\hline

\end{tabular}
  \caption{X-ray and hard X-ray spectral observations of 1E1724-3045
  in the globular cluster Terzan 2.  The models are defined as
  follows: PL=Power Law, BB=Single temperature blackbody, DBB=Disk
  Blackbody (Mitsuda et al. 1984), C=Comptt Comptonization model
  (Titarchuk 1994). $\alpha_{\rm PL}$ is the power law photon index.
  For the single temperature blackbody model: kT$_{\rm BB}$ is the
  Blackbody temperature, R$_{\rm BB}$ is the source radius in km.  For
  the multi-color blackbody model: kT$_{\rm in}$ is the temperature at
  the inner disk radius, R$_{\rm in}$ is the inner disk radius scaled
  at the source distance (7.7 kpc), $\theta$ is the inclination angle
  of the source. The column density (\nh) is given in units of
  $10^{22}$ \nhv. E refers to the energy range of the observation in
  keV. References are: 1) Mereghetti et al. (1995), 2) Zand (1992), 3)
  Verbunt et al. (1995), 4) Goldwurm et al. (1993), 5) Goldwurm et
  al. (1994), 6) Guainazzi et al. (1998), 7) Barret et al. (1998). The
  flux is the unabsorbed flux given in the energy band of the
  observation in units of $\times 10^{-10}$ \ergscm.}
\end{center}
\end{table*}

Thanks to SAX and RXTE, simultaneous X-ray to hard X-ray observations
have now been performed and have demonstrated that the 1-200 keV
spectrum of the source was indeed a more or less power law below $\sim
30$ keV, attenuated at high energies by an exponential cutoff at
$\sim$70 keV \cite{tz2:guainazzi98aa,tz2:barret98aa}.  This broad band
spectrum was interpreted as resulting from the Comptonization of soft
photons in a spherical scattering region of optical depth $\sim 3$ and
electron temperature $\sim 30$ keV. Furthermore, the latter
observations have suggested the presence of an additional soft
component below 2-3 keV, which could be fit either by a blackbody or a
muti-color disk blackbody model (See Table 1).  Beside the continuum,
no emission lines were found in the SAX/LECS-MECS data
\cite{tz2:guainazzi98aa}.

The absorbing column density (\nh) has been poorly determined:
depending on the model and the instrument, it varies from 0.5 to 4
$\times 10^{22}$ \nhv~from SAX to TTM. Table 1 summarizes the
X-ray/hard X-ray observations performed so far. Normalized to the
``classical'' 1-20 keV band, the fluxes listed in Table 1 indicate a
variability of about a factor of 3 between EXOSAT and TTM, SAX and
RXTE. A similar variability factor could be inferred in hard X-rays
\cite{tz2:goldwurm93:2ndcgro}. From the most recent measurements
performed by SAX and RXTE, the 1-200 keV luminosity lies in the range
$\sim 1.0$ to $1.5 \times 10^{37}$\ergs~(d=7.7 kpc).

Here we report on a short ASCA observation of \onee~in the 0.5--10 keV
range. Our observation provides an independent confimation of the
existence of a soft component in the source spectrum, yields the best
\nh~measurement obtained so far, puts stringent upper limit on disk
line emission, and finally by a combination with RXTE data enables us
to discuss on the origin of the soft and hard components.

\section{Spectral analysis}
The ASCA observation of \onee~were carried out on September 24-25th,
1995. Data were screened and analyzed using the standard procedure
described in the {\it The ASCA Guide to data reduction, version 2}
available at the ASCA Guest Observer
Facility\footnote{http://heasarc.gsfc.nasa.gov/docs/asca/abc/abc.html}.
In particular, data were rejected during times when the satellite was
traversing regions of low Cut-Off Rigidity (COR $\le 6$) and when the
Earth limb elevation angle was $\leq$10$^\circ$. The data were also
masked spatially in order to remove the outer ring of high background
as well as the calibration source events\footnote{see The ASCA Guide
to data reduction, p. 36}.  We also performed background rejection on
GIS data based on the rise-time of the signal.  The SIS screening
criteria for extracting spectral information were very similar to the
ones applied to the GIS: same minimum elevation angle and same minimum
rigidity cutoff.  Because of the possibility for contamination in the
SIS of fluorescence lines of oxygen from the Earth's atmosphere, data
selection was done on the bright-Earth angle, and only data above
40$^\circ$ (20$^\circ$) for the SIS0 (SIS1) were retained.  Finally,
only events with CCD grades 0, 2, 3, or 4 were used in further
analysis.

The position of the source from the image analysis is consistent with
with ROSAT and EINSTEIN positions
\cite{tz2:mereghetti95aa,grindlay84apjl}. In order to check for time
and spectral variability, we have produced background subtracted light
curves, as well as hardness-intenity and color-color diagrams. This
showed that the source is remarkably steady within our observation,
both in intensity and spectral shape.

The GIS spectrum was extracted from a circular region of radius
7$^\prime$ centered at the position of the source.  In order to
correctly subtract the contribution from the galactic plane diffuse
emission, we have extracted the background from multiple fields within
the FOV of the instrument.  The same procedure was applied to the SIS
data, except that a smaller region (3$^\prime$15$^{\prime\prime}$ of
radius) was used to extract the source spectrum. As mentioned above,
no spectral variations could be found within the observation, we have
therefore combined for each detector the whole data set to derive a
time averaged spectrum.  The spectral analysis has been restricted to
the 0.8 to 10 keV range for the GIS2 and GIS3 and 0.4, 0.5 to 10 keV
for SIS0 and SIS1.  In these energy bands, the source count rates are
7.20$\pm$0.03, 8.77$\pm$0.03, 8.40$\pm$0.03 and 7.04$\pm$0.03 count
s$^{ \rm -1}$ in the GIS2, GIS3, SIS0, SIS1 respectively. All fits
were done using XSPEC version 10.00.  We started by fitting the
detectors separately and then, merged them in a combined 4-detector
analysis.  Remarkable agreement is found between all the detectors and
we decide to keep the normalization equal for all the four
instruments.  Leaving the relative normalizations of the four data
sets as free parameters gives consistent results; the only noticeable
difference is that it leads to a reduction of the reduced $\chi^{2}$
values of about 0.2 (for more than 1300 degrees of freedom).
\subsection{Detection of the soft component}
Given previous observations of the source, we first fit the spectrum
with a single absorbed power law.  The power law fit is not good as
there is a clear low energy excess which is not accounted for in the
residuals.  We then tried to fit this excess with a blackbody (BB)
component.  This results in a significant improvement of the fits
($\Delta\chi^2 \ge 200$).  Taking advantage of the better spectral
resolution of the SIS, we have checked that adding such a soft
component is also required when fitting the SIS0 and SIS1 spectra
($\Delta\chi^2 \ge 150 $ for 408 degrees of freedom). We have also
tried to fit the soft component with a multi-color disk blackbody
(DBB, Mitsuda et al. 1984). This model provides an equally good
fit. The $\Delta\chi^2$ is less than 10 (1313 d.o.f) for the two
models. Other models such as a thermal Bremsstrahlung, a cutoff power
law, or a broken power law combined with a BB or a DBB could fit the
data as well but the flatness of the spectrum and the restricted
energy range of ASCA does not allow to put interesting constraints on
the fitted parameters.
\subsection{A Comptonization model to fit the ASCA data}
The source flux at the time of our observation was within 5\% of the
flux seen by SAX and within 30\% of the one observed by
RXTE. Likewise, the source spectrum was also very hard and a soft
component was also present; therefore it was legitimate to use the SAX
and RXTE models to try to fit the ASCA data as well.  This model is
made of the sum of either a BB or DBB component and the so-called {\em
comptt} model in XSPEC \cite{tita94apj}. This model is an improvement
over the former {\em compst} model as the theory is extended to
include relativistics effects, and to work for both the optically thin
and thick cases \cite{tita94apj}.  The free parameters of the model
are: the temperature of the seed photons (\ktw), assumed to follow a
Wien law, the optical depth of the scattering region ($\tau$), the
electron temperature (\kte), and a parameter defining the geometry (a
disk or a sphere).  In the following, we consider the spherical case.
Obviously in the absence of significant cutoffs in the spectrum, the
electron temperature cannot be easily constrained by the ASCA data.
Therefore, to begin with, we have set \ktw, $\tau$, \kte~to their best
fit values listed in Guainazzi et al.  \cite*{tz2:guainazzi98aa}; i.e.
1. keV, 3., and 30 keV respectively (see Table 1).  We thus fit the
normalization of the {\em comptt} model, and the soft excess modeled
either by a BB or a DBB (a fit is not possible when such a component
is not added).  The results of the fit are listed in Table 2.  The
lowest reduced $\chi^2$ value is obtained when the soft component is
fit by a BB. A comparison with the values listed in Table 1 indicates
that our best fit parameters are consistent with SAX for the
\comptt+BB model, but differs in the inner disk temperature for the
\comptt+DBB model.  Assuming a disk geometry instead of a spherical
one leads to the same conclusion. As a further test of the validity of
the latter model, we have let \ktw~and $\tau$ as free parameters of
the fit in the
\comptt+BB model.  The parameters so recovered \nh=$0.94\pm0.03 \times
10^{22}$ \nhv, $\tau=3.2^{+0.4}_{-0.7}$, \ktw=$1.1^{+0.1}_{-0.2}$ keV
while the parameters of the BB component remained unchanged.
\subsection{Joint fit of the ASCA and RXTE data}
\begin{figure}[!t]
\centerline{\psfig{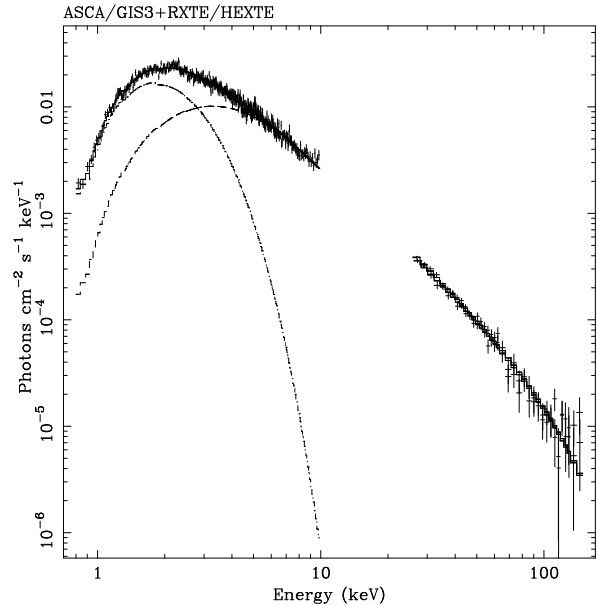}}
\caption{The unfolded ASCA/GIS3 and RXTE/HEXTE data 
(Barret et al. 1998) for the \comptt+BB model with parameters listed
in Table 2. The blackbody component is indicated by a dashed line.}
\label{gis3_hexte}
\end{figure}

\begin{figure}[!h]
\centerline{\psfig{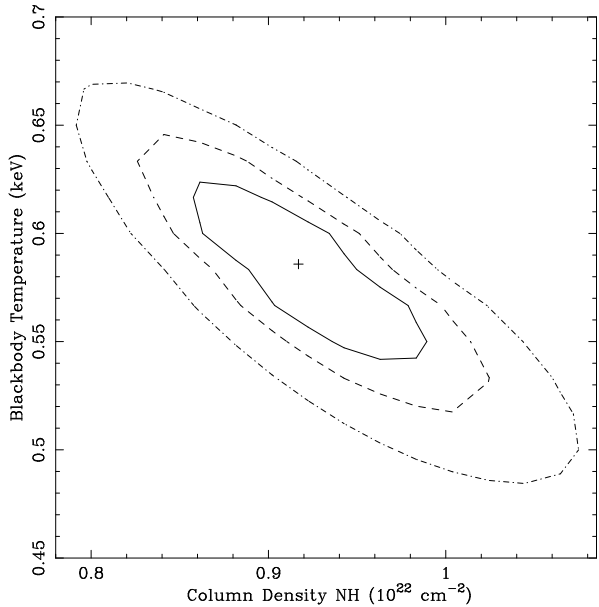}}
\caption{Contours of confidence level for \nh~and the blackbody 
temperature for the {\it comptt}+BB model fitting the ASCA/GIS3 and
RXTE/HEXTE data.  The contour levels shown are respectively: 68\% ,
90\%, 99\% confidence level.  The best fit is marked by a small cross
on the plot.}
\label{contours}
\end{figure}

We have shown above that the ASCA spectra alone could be fit by the
SAX and RXTE model. As a final test, we have tried to fit jointly the
ASCA data and the non simultaneous RXTE/HEXTE hard X-ray
spectrum. \onee~was observed by RXTE about one year after ASCA (from
November 4th to 8th for a total of $\sim 100$ ksec, see Barret et
al. 1998 for details about the observation). The HEXTE spectrum is a
time-averaged spectrum over the entire observation, and is made of the
cluster A and B spectra. We do not include the PCA data, as the PCA
has no sensitivity below 3 keV (i.e. \nh~and
\ktw~will be better constrained with ASCA alone), and the power law part of
the ASCA spectrum is sufficient to determine reliably the parameter
$\tau$ of the \comptt~model (see below). For the fit, we have let the
relative normalizations of the ASCA and two HEXTE cluster spectra be a
free parameter of the fit, to account for the fact that the
observations are not simultaneous as well as relative calibration
uncertainties between the instruments. In order not to give too much
weight to the ASCA data in the fit, we used only the GIS3 spectrum.
We used the
\comptt~model assuming a spherical geometry. The results of the fit
are listed in Table 2. As can be seen, an excellent fit is
obtained. Unfortunately, once again we are unable to distinguish
between the BB and DBB model to fit the soft component.  In
Fig. \ref{gis3_hexte}, we show the unfolded broad-band ASCA/GIS3 and
RXTE/HEXTE spectrum of the source associated with the \comptt+BB best
fit. In Figure \ref{contours}, we show the allowed grid of variations
for the \nh~and the blackbody temperature for that model.
\subsection{No emission lines}
No visible emission lines are present in the source spectrum.  We have
set an upper limit on the presence of an Iron line by adding a
gaussian line centered at 6.4 keV to the model described above.
For a line width of 0.1 keV ($\sigma$) , the upper limit on its
equivalent width is $\sim 30$ eV. This limit increases up to $\sim 80$
eV when $\sigma=0.5$ keV (both values are given at the 90\% confidence
level).
\section{Discussion}
Our observation confirmed the existence of a soft component in the
X-ray spectrum of \onee~at a luminosity around 5.7 $\times 10^{36}$
\ergs~(0.5-10.0 keV, 7.7 kpc). The non-detection of this soft
component in the EXOSAT and TTM data could be explained by a lack of
sensitivity below 2 keV.  If this soft component is a multi-color disk
blackbody, then our estimate of \rin $\sqrt{\cos\theta}$ is consistent
with the low value generally observed from neutron star systems, and
much smaller than the values derived from black hole candidates. On
the other hand, this is not the case for \ktin~which is much smaller
than the 1.5 keV neutron star value, and close to the black hole
values \cite{tanaka95review}. As for \rin, as pointed out by Guainazzi
et al. \cite*{tz2:guainazzi98aa}, in order for this radius to
accommodate a neutron star radius, a large inclination $\theta \ga
70^{0}$ is required, in contradiction with the fact that no sizable
orbital modulations have been observed from the source
\cite{tz2:olive98aa}.

For the \comptt+BB model, the ratio of the luminosities 0.1-200 of the
BB (\lbb) component versus the \comptt~one (\lcomptt) is
0.13-0.15. The ratio of \ldbb/\lcomptt~is 0.23-0.40 for the
\comptt+DBB model. The origin of the soft component is
unclear. Obviously it could come from the neutron star surface itself
or from an optically thick boundary layer near the neutron star. In
this scenario, the Comptonized component would be generated somewhere
in a disk corona. The blackbody radius of $\sim 10$ km derived from
the \comptt+BB clearly argues in favor of this scenario. Another
possibility is that the soft component originates from the accretion
disk, while the harder X-rays are generated in the boundary layer
\cite{tz2:guainazzi98aa}. Sunyaev and Shakura \cite*{sunyaev86sal}
predicted that the ratio between the disk and the boundary layer
luminosities could be as low as 0.45, if the disk ends at the
marginally stable orbit (which is larger than the neutron star
radius). As noticed by Guainazzi et al. \cite*{tz2:guainazzi98aa}, the
properties of the high frequency quasi-periodic oscillations recently
discovered in LMXBs \cite{vdklis97review} indicates that these systems
do not rotate at their break-up periods, and further do not have
accretion disks that extend beyond the last marginally stable
orbit. In other words, this means that the predicted ratio discussed
above could be even smaller than $\sim 0.45$. Thus the low ratio of
the soft/hard flux (i.e. BB or DBB vs. Comptonized flux) would favor
the picture in which the weak soft component comes from the accretion
disk and the Comptonized component originates from a hot and optically
thin boundary layer. This in turn would support theoretical models
predicting hard X-ray emission from such a boundary layer
(e.g. Kluzniak and Wilson 1991; Walker 1992).

\begin{table*}[t]
\scriptsize
\begin{center}
\hspace*{-0.7cm}\begin{tabular}{llllllllllccc}
\hline
Set & Model &\nh&\kte& $\tau$ & \ktw & \ktbb&\rbb & \ktin & \rin$\sqrt{\cos\theta}$ & $\chi^2$ & \ft 
& \fbb \\
\hline
1 & C+BB & $0.97^{+0.04}_{-0.04}$ & 30.0 (fixed) & 3.0 (fixed) & 1.0
(fixed) & $0.54^{+0.02}_{-0.02}$ & $10.8^{+0.2}_{-0.2}$ & & & 1.35 &
16.4 & 1.9 \\
\hline
\ldots & C+DBB & $1.28^{+0.01}_{-0.01}$ & 30.0 (fixed) & 3.0 (fixed) & 
1.0 (fixed) & & & $0.76^{+0.02}_{-0.02}$ & $5.2^{+0.2}_{-0.2}$ & 1.40
& 17.5 & 3.3 \\
\hline
2 & C+BB & $0.91^{+0.08}_{-0.07}$ & $27.0^{+11.0}_{-5.0}$ & 
$\tau=3.2^{+0.7}_{-0.9}$ & $1.15^{+0.13}_{-0.13}$ & $0.58^{+0.05}_{-0.05}$ &
$10.0^{+1.2}_{-1.2}$ & \ldots & \ldots & 1.09 & 16.0 & 2.1\\ 
\hline
\ldots & C+DBB & $1.21^{+0.07}_{-0.05}$ & $27.9^{+12.0}_{-5.0}$ & 
$\tau=3.0^{+0.8}_{-0.6}$ & $1.23^{+0.17}_{-0.23}$ & \ldots & \ldots &
$0.58^{+0.05}_{-0.05}$ & $2.45^{+0.5}_{-0.5}$ & 1.09 & 17.8 & 5.0 \\
\hline
\hline
\end{tabular}
\caption{Best fit results for the \comptt~ + BB or DBB model. The first 
data set corresponds to the ASCA GIS+SIS data. The parameters of the
\comptt~model have been fixed to those observed by SAX and RXTE (see Table 1). 
The second data set corresponds to the combination of the ASCA GIS3
and HEXTE spectra. The number of degrees of freedom for the first data
set is 1314, whereas it is 537 for the second one. The data are not
simultaneous so the relative normalization between the ASCA and HEXTE
spectra has been left as a free parameter. For the \comptt~model, a
spherical geometry has been assumed. \ft~is the unabsorbed flux in the
0.1-200 keV band, \fbb~is the flux carried of by the soft component in
the same energy band. They are given in units of $10^{-10}$ \ergscm.}
\end{center}
\end{table*}

Our observation provides the most accurate \nh~estimate obtained so
far (see Fig. \ref{contours}). Our values are consistent with the one
derived by SAX \cite{tz2:guainazzi98aa}. Following Predehl and Schmitt
\cite*{predehl95aa}, the visual extinction is related to the X-ray
absorbing column density as \nh=$A_{\rm V} \times 10^{21}=R\times
E(B-V) \times 10^{21}$ \nhv.  Taking E(B-V)=1.54
\cite{tz2:ortolani97aa}, we obtain \nh=$0.9$ and $1.0 \times 10^{22}$
\nhv~for R=3.1 and 3.6 respectively.  The ``optical'' value is
remarkably close to the value derived with the \comptt+BB model (see
Table 2). It is smaller by than the value derived for the \comptt+DBB
model (the systematic difference in the
\nh~value between the BB and DBB models is due to the fact that the
DBB is steeper than the BB model at low energies). In any case, the
difference between the observed and expected \nh, and the consistency
between the SAX, RXTE values suggest that the source does not show
intrinsic (and variable) absorption, as one might have suspected from
previous measurements (see Table 1). The larger \nh~observed by TTM
and at a lower degree by EXOSAT might result from an overestimate of
the photon index for a power law fit.  We now know from SAX, RXTE and
the present data (all of which were in a very similar spectral state)
that the photon index is less than 2 for a power law fit. Our results,
which show agreement between the ``optical'' \nh~and the X-ray
\nh~values, therefore eliminate the need for internal absorption in
the source and therefore, any dependence upon the metallicity of
Terzan 2 (which happens to be metal rich).


Consistent with the SAX/LECS-MECS observations, ASCA failed to detect
any emission lines from \onee.  For instance, an iron line and a broad
absorption feature above 10 keV were detected from the X-ray burster
4U1608-522 by GINGA, and interpreted as due to absorption/reflection
of a power law incident spectrum on a relatively cold medium
\cite{1608:yoshida93pasj}.  Therefore the lack of an iron line in our
data is consistent with the absence of a reflected component the
broad-band SAX and RXTE spectra of
\onee~\cite{tz2:guainazzi98aa,tz2:barret98aa}. It is also consistent
with the emission line models of Ko and Kallman \cite*{ko94apj} for
this relatively low luminosity source (compared to Sco X-1 which does
show Fe line emission), particularly if the disk in
\onee~is illuminated at relatively grazing angles from the central
x-ray source.

\section{Conclusions}
We observed \onee~in its hard state, during which it emits hard
X-rays.  The weak variability of the source in X-rays suggests that it
spends most of its time in such a hard state with a hard power law
type spectrum in X-rays (0.5-10 keV).  The soft excess detected at low
energies is more likely originating from the accretion disk. As for
the hard component, it more likely originates from the comptonization
of soft photons in a hot boundary layer. The \nh~value we have
measured is consistent with the value expected from the cluster
redenning. The consistency of this result with recent
\nh~measurements by SAX and RXTE suggests that the source does not
display intrinsic (and variable) absorption. The lack of iron line
emission is consistent with the absence of a reflected component in
the broad band spectrum of the source.
\section{Acknowledgments}
Part of this work was supported by NASA Grant (to JEG) NAG5-2763. This
research has made use of data obtained through the High Energy
Astrophysics Science Archive Research Center operated by the
NASA/Goddard Space Flight Center.

The authors are grateful to L. Titarchuk for useful discussions, and
to A. Parmar and M. Guainazzi for providing us with the SAX results
prior to publication.

\end{document}